\documentclass[]{spie}  

\usepackage{amsmath,amsfonts,amssymb}
\usepackage{graphicx}
\usepackage[colorlinks=true, allcolors=blue]{hyperref}

\title{The AI Security Pyramid of Pain}

\author{Chris M. Ward}
\author{Josh Harguess}
\author{Julia Tao}
\author{Daniel Christman}
\author{Paul Spicer}
\author{Mike Tan}

\affil{Cranium AI \\ Short Hills, NJ, USA}

\authorinfo{Further author information: (Send correspondence to Chris M. Ward)\\Chris M. Ward: E-mail: cward@cranium.ai}

\begin{document} 
\maketitle

\begin{abstract}

With resource constraints being a reality for many organizations, effective security for Artificial Intelligence (AI) systems has become a critical focus. To address this challenge, we introduce the AI Security Pyramid of Pain, a framework that adapts the cybersecurity Pyramid of Pain to categorize and prioritize AI-specific threats. This framework provides a structured approach to understanding and addressing various levels of AI threats.

Starting at the base, the pyramid emphasizes Data Integrity, which is essential for the accuracy and reliability of datasets and AI models, including their weights and parameters. Ensuring data integrity is crucial, as it underpins the effectiveness of all AI-driven decisions and operations. The next level, AI System Performance, focuses on MLOps-driven metrics such as model drift, accuracy, and false positive rates. These metrics are crucial for detecting potential security breaches, allowing for early intervention and maintenance of AI system integrity. Advancing further, the pyramid addresses the threat posed by Adversarial Tools, identifying and neutralizing tools used by adversaries to target AI systems. This layer is key to staying ahead of evolving attack methodologies. At the Adversarial Input layer, the framework addresses the detection and mitigation of inputs designed to deceive or exploit AI models. This includes techniques like adversarial patterns and prompt injection attacks, which are increasingly used in sophisticated attacks on AI systems. Data Provenance is the next critical layer, ensuring the authenticity and lineage of data and models. This layer is pivotal in preventing the use of compromised or biased data in AI systems. At the apex is the tactics, techniques, and procedures (TTPs) layer, dealing with the most complex and challenging aspects of AI security. This involves a deep understanding and strategic approach to counter advanced AI-targeted attacks, requiring comprehensive knowledge and planning.

This paper provides an overview of each layer, discussing their significance and interplay within AI system security. The AI Security Pyramid of Pain offers a strategic approach for organizations to navigate and mitigate the multifaceted threats in AI security effectively.
\end{abstract}

\keywords{AI Security Framework, Tactics, Techniques, and Procedures (TTPs), AI Red Teaming}

\section{INTRODUCTION}
\label{sec:intro}  

The landscape of cybersecurity has evolved significantly with the advent of Artificial Intelligence (AI) technologies. As AI systems become increasingly integral to various sectors, the need for robust security frameworks tailored to these systems is paramount. This paper introduces the ``AI Security Pyramid of Pain'', a novel framework designed to categorize and prioritize threats specific to AI systems and organization goals.

The concept of the Pyramid of Pain in cybersecurity was first introduced by David Bianco \cite{bianco2013pyramid}. His framework provides a hierarchy of cyber threat indicators, emphasizing the importance of addressing more sophisticated attack vectors. Inspired by this concept, the AI Security Pyramid of Pain adapts these principles to the realm of AI security, addressing the unique challenges and threats in this field.

Moreover, the concept of a structured approach to cybersecurity is not new. The Cyber Kill Chain framework \cite{hutchins2011intelligence}, developed by Hutchins, Cloppert, and Amin, describes the stages of a cyber attack, offering insights into the process of intrusion and defense. These foundational concepts have significantly influenced the development of specialized security strategies in various domains, including AI. Similarly, the need for specialized security in AI has been highlighted in recent literature. Papernot, McDaniel, and Goodfellow \cite{papernot2016transferability} discuss the unique vulnerabilities inherent in AI systems, particularly in the context of adversarial attacks, underscoring the need for frameworks like the AI Security Pyramid of Pain.

We build upon these foundational works, extending the pyramid model to the realm of AI. The importance of early detection and response in cybersecurity, a principle well-documented by researchers like Liao \texttt{et al.} \cite{liao2013deep}, is foundational to the base level of the AI Security Pyramid of Pain, which prioritizes fundamental AI system monitoring. As AI technologies advance, the sophistication of potential attacks also increases. The work of Biggio and Roli \cite{biggio2018wild} on data integrity and secure AI development practices underscores the necessity of these elements in the middle tiers of the pyramid. These practices are crucial in safeguarding AI systems from more advanced threats.

At the top of the pyramid lies the challenge of understanding and mitigating advanced adversarial tactics, techniques, and procedures (TTPs). The concept of TTP-based hunting in cybersecurity, as discussed by Daszczyszak \texttt{et al.} \cite{daszczyszak2019ttp}, emphasizes the importance of understanding and mitigating advanced adversarial tactics, techniques, and procedures in AI security. This approach aligns with the apex of the AI Security Pyramid of Pain, where strategic and informed responses to cyber threats are crucial.

The AI Security Pyramid of Pain provides a structured, hierarchical framework for addressing the spectrum of threats faced by AI systems. It is a response to the growing need for specialized security measures in the field of AI, offering a practical tool for AI security professionals within the resource constraints of modern organizations.

\section{BACKGROUND}
\label{sec:background}  

The intersection of Artificial Intelligence (AI) and cybersecurity has become a critical area of concern in recent years. As AI systems are increasingly deployed in various domains, their security vulnerabilities have become apparent, necessitating a specialized approach to AI security.

\subsection{Evolution of AI Security Concerns}
The evolution of AI security concerns can be traced back to the early days of AI development. Initially, the focus was primarily on improving performance and capabilities. However, as AI systems became more pervasive, the security implications of these systems began to surface. Kurakin \texttt{et al.} \cite{kurakin2016adversarial} demonstrated how AI models, specifically deep neural networks, could be easily misled by adversarial examples in real-world scenarios. This revelation marked a significant shift in the AI research community, leading to a greater focus on the security aspects of AI.

\subsection{Adversarial Attacks and Defenses}
Adversarial attacks involve manipulating input data to AI systems in a way that causes them to malfunction. This can have serious implications, especially in critical applications like autonomous vehicles or healthcare. Akhtar and Mian \cite{akhtar2018threat} provided a comprehensive overview of adversarial attacks in the context of deep learning. Since then, a significant body of research has emerged, focusing on both developing adversarial attack techniques and devising defenses against them. For instance, Tramèr \texttt{et al.} \cite{tramer2017ensemble} explored ensemble adversarial training as a defense mechanism.

\subsection{AI in Cybersecurity Applications}
Conversely, AI has also been leveraged to enhance cybersecurity. Machine learning models are increasingly used for threat detection and response. Apruzzese \texttt{et al.} \cite{apruzzese2018effectiveness} discussed the effectiveness of machine learning in cybersecurity, particularly in anomaly detection. However, the use of AI in security also introduces new vulnerabilities, as attackers can target the AI models themselves, as noted by Papernot \texttt{et al.} \cite{papernot2016towards}.

\subsection{The Need for a Structured Security Framework}
The complexity and evolving nature of AI security threats necessitate a structured approach to AI security. The AI Security Pyramid of Pain, inspired by the Cybersecurity Pyramid of Pain \cite{hutchins2011intelligence}, aims to address this need by providing a hierarchical framework for understanding and mitigating AI security threats. This framework is particularly relevant in the context of the increasing sophistication of AI systems and the corresponding advancement of adversarial techniques, as discussed by Yuan \texttt{et al.} \cite{yuan2019adversarial}.

\subsection{Emerging Trends in AI Security}
Recent years have seen emerging trends in AI security, such as the use of federated learning for privacy-preserving AI models, as explored by Konečný et al. \cite{konecny2016federated}. Additionally, the rise of quantum computing presents new challenges and opportunities in AI security, as noted by Mosca \texttt{et al.} \cite{mosca2019cybersecurity}. These emerging trends underscore the dynamic nature of the field and the need for adaptive security frameworks.

\subsection{Challenges in AI Security}
Despite advancements in AI security, several challenges remain. 

The field of AI security faces several pressing challenges. AI models are susceptible to advanced and sophisticated hacking techniques.\cite{oseni2020security}. This vulnerability necessitates ongoing research in adversarial AI to develop robust models capable of withstanding diverse adversarial scenarios. Understanding the intentions and capabilities of adversaries is crucial to formulate adaptive defense strategies for AI applications.

Another significant challenge arises in the context of cloud or fog computing-based AI applications. Pakmehr \texttt{et al.} emphasize the importance of securing underlying cloud or fog services. Attacks on these services can have a drastic impact on AI applications, leading to serious impairments. The paper explores various threats and attack possibilities specific to cloud and fog platforms, highlighting the distinct security requirements essential for AI applications in these environments \cite{pakmehr2020security}.

AI security is challenged not only by technical vulnerabilities but also by the lack of regulation, leading to potential cyber arms races. Taddeo and Floridi\cite{taddeo2018regulate} emphasize the necessity of regulating AI to prevent such escalation, especially in cybersecurity. The rapid advancement of AI technologies without adequate ethical guidelines and regulatory frameworks poses significant risks. Moreover, international cooperation is crucial in establishing responsible and safe AI development and use.

The background of AI security is characterized by a growing awareness of the vulnerabilities inherent in AI systems and the corresponding need for robust security measures. The AI Security Pyramid of Pain represents a response to these challenges, offering a structured approach to navigating the complex landscape of AI security.

\begin{figure}[h]
  \centering
  \includegraphics[width=0.80\textwidth]{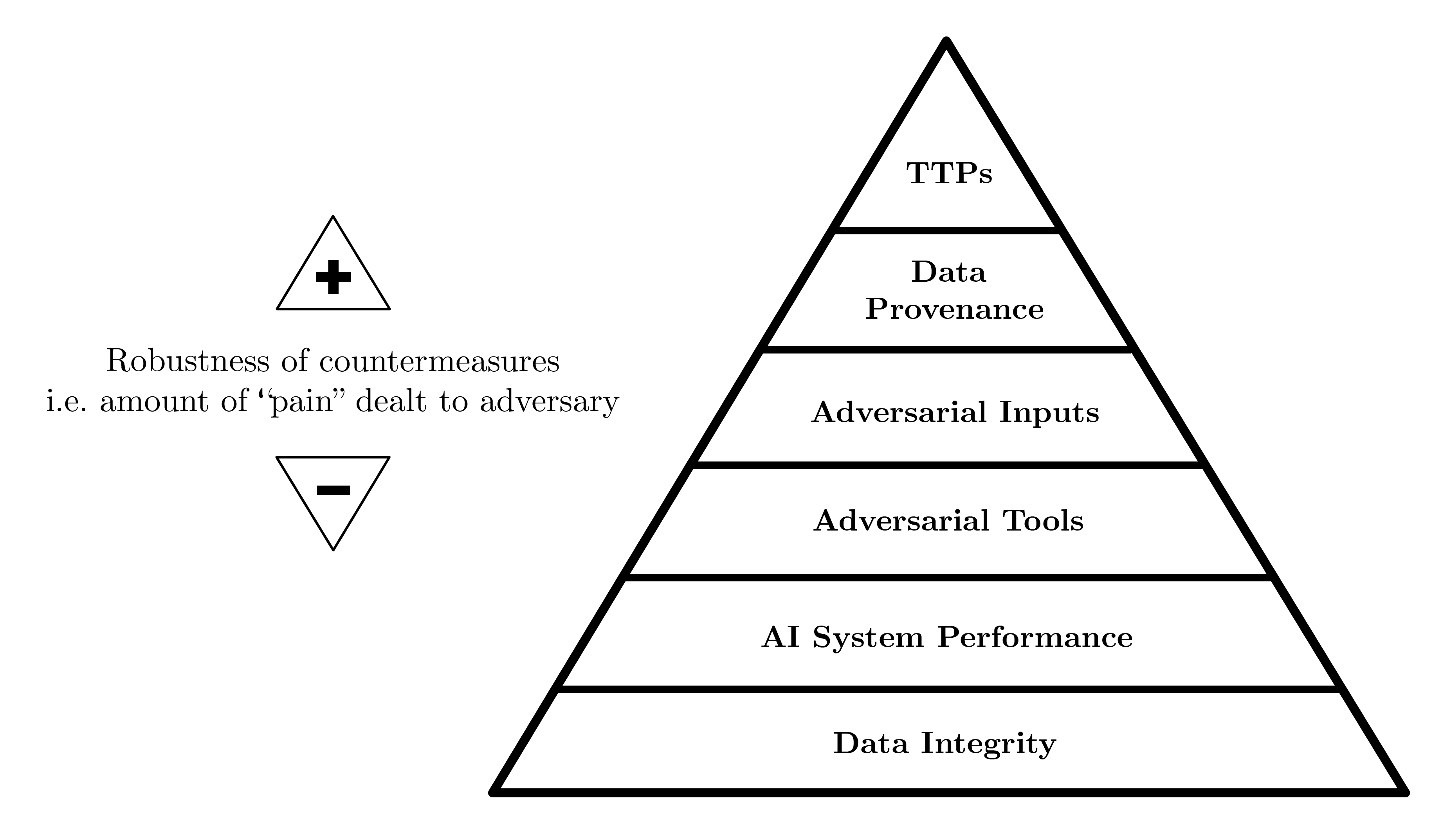}
  \caption{The AI Security
Pyramid of Pain \textsuperscript{\textcircled{\tiny R}}}
  \label{fig:AI Pyramid of Pain}
\end{figure}

\section{AI SECURITY PYRAMID OF PAIN}
\label{sec:background}  

The ``The AI Security Pyramid of Pain'' (Figure \ref{fig:AI Pyramid of Pain}) is a conceptual framework that illustrates the varying levels of difficulty for adversaries when they are confronted with different types of AI Security countermeasures. 

This section explores each layer of the AI Security Pyramid of Pain, discussing the theoretical foundations and practical implications.

\subsection{Data Integrity}

Data Integrity forms the foundation of the AI Security Pyramid. It involves ensuring the accuracy, consistency, and reliability of data throughout its lifecycle. This is crucial for AI systems, as they rely heavily on data for training, learning, and decision-making. The integrity of data affects everything from model training to real-world AI application performance.

Ensuring data integrity involves several key aspects:

\begin{itemize}
    \item \textbf{Accuracy and Consistency}: The data used in AI systems must be accurate and consistent. Inaccuracies or inconsistencies in training data can lead to flawed learning and incorrect outcomes. Techniques such as data validation and cleansing are essential to maintain the accuracy and consistency of data.

    \item \textbf{Reliability}: The reliability of data refers to its quality and trustworthiness over time. Reliable data sources and consistent data collection processes are vital to ensure that the data remains valid for AI applications.

    \item \textbf{Data Protection}: Protecting data from unauthorized access and tampering is crucial. This involves implementing security measures such as encryption, access controls, and secure data storage solutions.

    \item \textbf{Checksums and Hash Functions}: Checksums and cryptographic hash functions are used to verify the integrity of data. They help in detecting alterations or corruptions in data, ensuring that the data used in AI systems has not been tampered with.

    \item \textbf{Audit Trails}: Maintaining audit trails for data helps in tracking its origin, modifications, and the various processes it has undergone. This is particularly important in scenarios where data provenance and compliance are critical.
\end{itemize}

In summary, data integrity is a multifaceted concept that encompasses various practices and techniques to ensure that the data used in AI systems is accurate, consistent, reliable, and secure. It forms the bedrock upon which effective and trustworthy AI systems are built.

\subsection{AI System Performance}

The second layer of the AI Security Pyramid of Pain, System Performance, is pivotal in ensuring the efficiency and effectiveness of AI systems. This layer emphasizes the importance of MLOps-driven metrics and performance monitoring, which are crucial for maintaining the reliability and robustness of AI systems.

Key aspects of System Performance include:

\begin{itemize}
    \item \textbf{Model Drift}: Over time, AI models can drift from their initial performance due to changes in the underlying data or environment. Monitoring for model drift is essential to ensure that the AI system continues to perform as expected. Techniques like retraining the model with new data or adjusting the model parameters are often used to counteract drift.



    \item \textbf{Inference Performance Metrics and Evaluation}: To gain a comprehensive understanding of the strengths and weaknesses of algorithms and their relevance to specific domains or tasks, evaluating their performance with suitable metrics is essential. Various metrics like accuracy, precision curves, precision-recall curves, and receiver operating characteristic (ROC) curves are commonly utilized in research. It's equally crucial to select an appropriate evaluation method that determines whether a detection is true or false. The choice of both the metric and the evaluation method is vital as it can lead to either a positive or negative bias in the evaluation process, potentially favoring or disadvantaging certain algorithms or tasks \cite{parameswaran2016evaluation}. Regular evaluation against benchmark datasets is necessary to ensure that the model maintains high accuracy levels. This may involve techniques such as cross-validation and performance tuning.

    \item \textbf{Computational Efficiency}: The computational efficiency of AI models affects their scalability and deployment feasibility. Optimizing models for faster processing without compromising accuracy is a key challenge, especially for applications requiring real-time analysis.

    \item \textbf{MLOps Metrics and Monitoring Tools}: MLOps (Machine Learning Operations) metrics and monitoring tools are essential for continuous evaluation and improvement of AI systems. These tools help in tracking performance metrics, identifying issues, and facilitating the deployment of updated models.
\end{itemize}

System Performance is a dynamic layer that requires continuous attention and adaptation. By focusing on these key aspects, organizations can ensure that their AI systems not only perform efficiently but also remain robust and reliable over time.

\subsection{Adversarial Tools}

The Adversarial Tools layer in the AI Security Pyramid of Pain is dedicated to understanding and neutralizing the tools and methods used by adversaries to create adversarial examples or inputs that can deceive AI models. This might involve algorithms that can generate inputs to cause a machine learning model to misclassify, make errors, or reveal sensitive data.

Key aspects of the Adversarial Tools layer include:

\begin{itemize}

     \item \textbf{Rapid Development} As in traditional cybersecurity, adversarial tools targeting AI systems are constantly being developed and refined. Attackers often customize their tools to target specific vulnerabilities or to evade detection.
    AI-accelerated development of adversarial tooling has  made the tracking of such tools—as previously done in cybersecurity—increasingly difficult to manage.

      \item \textbf{AI Augmentation}  Many of the tools used in adversarial AI attacks leverage AI to formulate highly effective attacks and adversarial artifacts. This may include tools that generate adversarial image perturbations, adversarial prompts, or poisoned data. 

    \item \textbf{Adversarial Tools for Defense} Adversarial methods can be leveraged defensively, by utilizing openly available tools like IBM Adversarial Robustness Toolkit\cite{nicolae2018adversarial}. These tools can be used to test for robustness against adversarial inputs, or create adversarial samples to be used in adversarial training.  
      
    \item \textbf{Adversarial Training}: Adversarial training involves exposing the AI model to adversarial examples during the training phase. This technique aims to make the model more robust by learning to correctly classify these manipulated inputs. It's a proactive approach to improve the model's resilience against such attacks. Developments in this field have been comprehensively reviewed by Bai \texttt{et al.}, offering a new taxonomy and highlighting ongoing challenges in adversarial training \cite{bai2020advances}.

\end{itemize}

The Adversarial Tools layer is dynamic, requiring continuous updates and adaptations to stay effective against evolving threats. By focusing on these key aspects, organizations can build a robust defense against the sophisticated tools and methods used by adversaries in targeting AI systems.

\subsection{Adversarial Input Detection}

Adversarial Input Detection is a critical layer in the AI Security Pyramid of Pain, addressing the vulnerabilities of AI systems to adversarial attacks. In these attacks, inputs are deliberately manipulated to cause the AI system to make errors or produce incorrect outputs. This layer focuses on identifying and mitigating such threats through various techniques and strategies.

Key aspects of Adversarial Input Detection include:

\begin{itemize}
    \item \textbf{Understanding Adversarial Attacks}: The first step in defending against adversarial attacks is understanding how they work. This involves studying different types of adversarial inputs and the methods used to generate them. Research in this area provides insights into the vulnerabilities of AI models and helps in developing effective countermeasures. 
    \cite{vassilev2024adversarial}.

    \item \textbf{Anomaly Detection Algorithms}: Anomaly detection plays a significant role in identifying unusual patterns or changes in input data that could signify an adversarial attack. These algorithms are designed to flag anomalies for further investigation, thereby providing an additional layer of defense \cite{parameswaran2016evaluation}.

    \item \textbf{Input Validation and Sanitization}: Rigorous input validation and sanitization can prevent many adversarial attacks by ensuring that the input data conforms to expected norms and is free from malicious manipulations. This is particularly important in systems that interact with external data sources.

    \item \textbf{Regular Model Evaluation and Updates}: Continuous evaluation and timely updates of AI models are essential to maintain their defense against evolving adversarial techniques. This involves regularly testing the models against new adversarial examples and updating them to address any identified vulnerabilities. \cite{holt2021baseline}

    \item \textbf{Adversarial Pattern Detection}: Incorporating adversarial pattern detection techniques is crucial for identifying and mitigating physical adversarial attacks, such as those involving adversarial patches that can deceive object detectors.\cite{jutras2022detecting} Utilizing datasets like APRICOT\cite{braunegg2020apricot}, which focuses on physical adversarial attacks on object detection, can aid in benchmarking the effectiveness of defenses against such threats.

\end{itemize}

Adversarial Input Detection is an ongoing process that requires vigilance and adaptation to new threats. By focusing on these key aspects, organizations can enhance the security and reliability of their AI systems against adversarial attacks.

\subsection{Data Provenance}

Data Provenance is a crucial layer in the AI Security Pyramid of Pain, focusing on ensuring the authenticity and lineage of data and models. This layer is instrumental in preventing the use of compromised or biased data in AI systems, which is vital for maintaining the integrity and reliability of AI-driven decisions.

Key aspects of Data Provenance include:

\begin{itemize}
    \item \textbf{Metadata Tagging}: Implementing metadata tagging involves attaching descriptive information to data sets and individual data elements. This metadata can include details about the origin of the data, the conditions under which it was collected, and any subsequent modifications. Metadata tagging is essential for tracing the history of data and understanding its context.

    \item \textbf{Blockchain for Data Tracking}: Utilizing blockchain technology for data tracking offers a secure and immutable record of data transactions. This can be particularly useful in scenarios where data provenance is critical, such as in healthcare or financial services. Blockchain ensures that once data is recorded, it cannot be altered retroactively, thereby enhancing trust and transparency.

    \item \textbf{Audit Trails and Logging}: Maintaining comprehensive audit trails and logs is essential for tracking data access and modifications. This not only helps in ensuring the integrity of the data but also aids in regulatory compliance and forensic analysis in the event of a security breach.

    \item \textbf{Data Source Verification}: Verifying the authenticity of data sources is a critical step in ensuring data provenance. This involves assessing the credibility and reliability of data providers and the methods used for data collection.

    \item \textbf{Version Control Systems}: Implementing version control systems for data and models allows for tracking changes over time. This is particularly important in collaborative environments where multiple stakeholders are involved in the data lifecycle. Version control provides a clear history of modifications, contributing to data transparency and accountability.
\end{itemize}

Data Provenance is a multifaceted layer that requires meticulous attention to detail and robust processes. By focusing on these key aspects, organizations can significantly enhance the authenticity and lineage of their data, thereby fortifying the overall security of their AI systems.

\subsection{Tactics, Techniques, and Procedures (TTPs)}

At the apex of the AI Security Pyramid of Pain are Tactics, Techniques, and Procedures (TTPs). This layer involves a deep understanding and strategic approach to counter advanced AI-targeted attacks. It requires comprehensive knowledge of AI security, including the latest developments in adversarial tactics, and the ability to develop and implement effective countermeasures.

Key aspects of the TTPs layer include:

\begin{itemize}
    \item \textbf{Threat Intelligence}: Keeping abreast of the latest adversarial techniques and tools in the AI landscape is essential. Threat intelligence involves gathering and analyzing information about existing and emerging threats. This knowledge helps in predicting and preparing for potential attacks.

    \item \textbf{Advanced Threat Modeling}: Developing sophisticated models of potential threats is crucial in this layer. This involves understanding the attacker's perspective and anticipating their moves. Advanced threat modeling helps in identifying potential vulnerabilities and preparing effective defenses.

    \item \textbf{Custom Defense Strategies}: Given the unique nature of AI systems, custom defense strategies are often required. These strategies are tailored to the specific characteristics and vulnerabilities of the AI system in question, ensuring a more effective defense against targeted attacks.

    \item \textbf{Continuous Security Training}: Keeping up with the rapidly evolving field of AI security demands continuous training and education. This includes staying informed about the latest research, attack methods, and defense strategies in AI security.

    \item \textbf{Incident Response and Recovery}: Preparing for and responding to security incidents is a critical component of TTPs. This involves having a well-defined incident response plan and the ability to quickly recover from attacks, thereby minimizing their impact.

    \item \textbf{Collaboration with AI Security Community}: Collaboration with the broader AI security community, including researchers, practitioners, and industry experts, is essential for staying ahead of threats. Sharing knowledge and experiences can lead to more robust and effective security practices. A good example of this knowledge sharing and collaboration is MITRE ATLAS\textsuperscript{\texttrademark} (Adversarial Threat Landscape for Artificial-Intelligence Systems) \cite{atlas}.
\end{itemize}

The TTPs layer represents the most sophisticated and dynamic aspect of the AI Security Pyramid of Pain. It requires a proactive and informed approach to effectively navigate and mitigate the complex landscape of AI-targeted threats.

\section{DISCUSSION AND FUTURE WORK}
\label{sec:background}  

\subsection{Discussion}

The AI Security Pyramid of Pain, as outlined in this paper, is designed with the primary goal of increasing the cost and complexity for adversaries in executing their attacks. By being a `hard target', AI systems can significantly reduce the success rate of adversarial attacks. This approach involves a spectrum of strategies, from proactive, behavior-based defenses at the top of the pyramid to more reactive, static defenses at the bottom.

\textbf{Top of the Pyramid - Proactive Defense:}
\begin{itemize}
    \item Proactive defense strategies, particularly those that are behavior-based, require more resources. This includes a combination of advanced technologies like machine learning and artificial intelligence, as well as skilled security personnel.
    \item Implementing threat indicators or detectors at this level is more challenging, but it significantly increases the difficulty for adversaries to adapt to the defenses.
\end{itemize}

\textbf{Bottom of the Pyramid - Reactive Defense:}
\begin{itemize}
    \item Reactive defense mechanisms are often static and signature-based. While these are easier to implement, using automated tests and software solutions, they may not be as resilient to changes in adversarial behaviors.
    \item Such solutions can quickly become obsolete as adversaries evolve their tactics, making continuous updates and adaptations necessary.
\end{itemize}

\subsection{Future Work}

Future research and development should focus on several key areas:

\begin{itemize}
    \item \textbf{Balancing Proactive and Reactive Approaches}: Strategies should be developed that effectively balance proactive and reactive defense mechanisms within the AI Security Pyramid of Pain.
    \item \textbf{Cost-Effective Proactive Solutions}: Cost-effective methods for implementing proactive defense strategies, particularly for organizations with limited resources, should be further explored.
    \item \textbf{Adaptive and Resilient Systems}: The adaptability and resilience of AI systems should be enhanced to handle rapidly evolving adversarial tactics and techniques.
    \item \textbf{Automated and Advanced Detection Tools}: Advancing the development of automated tools that can detect sophisticated attacks, reducing the reliance on manual intervention and expertise will be paramount for securing AI systems.
\end{itemize}





\section{CONCLUSION}
\label{sec:background}  
The AI Security Pyramid of Pain provides a strategic and structured approach to address the spectrum of threats in AI security. While it offers a comprehensive framework, the dynamic nature of AI technologies and the evolving threat landscape require continuous adaptation and innovation in security strategies. Future work in AI security is not only necessary but imperative for the safe and secure deployment of AI technologies. The ongoing development of automated tools, interdisciplinary collaboration, regulatory frameworks, and educational initiatives will be key to advancing the field of AI security and ensuring the resilience of AI systems against emerging threats.

\bibliography{main} 
\bibliographystyle{spiebib} 

\end{document}